\documentclass[12pt,american]{article}
\usepackage{lmodern}

\usepackage{babel}
\usepackage{soul}
\usepackage{xcolor}
\usepackage{geometry}
\usepackage{color}
\usepackage{verbatim}
\usepackage{amsmath}
\usepackage{amsfonts}
\usepackage{mathrsfs}
\usepackage{amsthm}
\usepackage{amssymb}
\usepackage{graphicx}
\usepackage{setspace}
\usepackage{bbm}
\usepackage{esint}
\usepackage{hhline}
\usepackage[colorlinks=true,citecolor=gray,linkcolor=black,urlcolor=gray]{hyperref}
\usepackage{dcolumn}
\usepackage{float}
\usepackage{booktabs}
\usepackage[normalem]{ulem}
\usepackage[longnamesfirst]{natbib}
\usepackage{multirow}
\usepackage{siunitx}
\usepackage{adjustbox}

\newcolumntype{d}{S[
    input-open-uncertainty=,
    input-close-uncertainty=,
    parse-numbers = false,
    table-align-text-pre=false,
    table-align-text-post=false
 ]}

\newcommand{\pdv}[2]{\frac{\partial #1}{\partial #2}}

\makeatletter
\usepackage[utf8]{inputenc}

\begin{document}

\title{Close Encounters of the LEO Kind: Spillovers and Resilience in Partially-Automated Traffic Systems
}

\author{Akhil Rao\footnote{Department of Economics, Middlebury College. E-mail: akhilr@middlebury.edu.} \thanks{I am grateful to Hugh Lewis for guidance on obtaining and working with the SOCRATES data, and to seminar participants at the United States Air Force Academy Department of Economics and Geosciences, Oregon State University Department of Economics, and Resources for the Future for helpful comments and feedback. All errors are my own.}}

\date{October 25, 2023 \\
Draft - Please do not cite
}

\maketitle
\thispagestyle{empty}

\begin{abstract}
\begin{onehalfspace}
Traffic systems are becoming increasingly automated. How will automated objects interact with non-automated objects? How will partially-automated systems handle large disruptions? Low-Earth orbit (LEO)---filled with thousands of automated and non-automated satellites and many more uncontrollable pieces of debris---offers a useful laboratory for these questions. I exploit the COSMOS-1408 (C1408) anti-satellite missile test of November 2021---a large and exogenous shock to the orbital environment---to study how an unexpected disruption affects a partially-automated traffic system. I use publicly-available close approach data, network theory, and an econometric analysis of the C1408 test to study the effect of close encounters with new fragments on the configuration of objects in orbit. I document spillover effects of close encounters with C1408 fragments, heterogeneity in impacts across operators, and changes in system-level resilience to new shocks. These results shed light on the nature of partially-automated traffic systems, and provide a basis for new models to anticipate and mitigate space traffic disruptions.
\end{onehalfspace}
\end{abstract}
\begin{onehalfspace}


 \end{onehalfspace}
\addtocounter{page}{-1}

\thispagestyle{empty} 

\newpage{}

\section{Introduction} 

As high-speed transportation systems become increasingly automated, understanding interactions between payloads and the risks of cascading traffic disruptions is increasingly important. The potential for catastrophic collisions between objects moving at high speeds, some using automated decision systems and others relying on human input, makes it important to understand causes and consequences of unsafe traffic patterns. It is as-yet unclear how interactions between payloads will affect overall traffic patterns and the risk of low-probability high-impact failures like collision cascades. Low-Earth orbit (LEO), where thousands of satellites with and without automated maneuvering have recurrent, publicly-observable, high-speed close approaches with each other and uncontrollable debris objects, offers a unique setting to study these issues. \\

How will partially-automated traffic systems handle disruptions? Will close encounters with disruptive uncontrollable objects crowd in or crowd out close encounters between other objects? How will automated maneuvering affect the overall traffic system's resilience to collision cascades? This paper addresses these questions by using the COSMOS-1408 (C1408) missile test in 2021---a large and unexpected disruption to space traffic. I use publicly-available data on predicted close approaches (``conjunction alerts'') for nearly all objects in LEO, network theory, and econometric analysis to investigate the effect of close encounters with C1408 fragments on the configuration of objects in orbit. I find that close approaches with C1408 fragments crowded out close approaches with non-C1408 objects, with larger effects for objects ``closer'' to C1408. I also find that automated maneuvers reduced the risk of collision cascades, while increasing the strength of potential strategic interactions between payloads. \\

C1408 was a defunct Russian military satellite launched on 16 September 1982 for signals intelligence collection. It operated for roughly two years, after which it was left in orbit as a derelict object. Russia conducted an anti-satellite missile (ASAT) test against C1408 on 15 November 2021, using a ground-based missile to destroy the satellite and create a large debris field in LEO.\footnote{There is not clear consensus as to \textit{why} Russia conducted this test at that particular time and using that particular object. U.S. Space Command has alleged that the test was motivated by an interest in actively denying access to and use of space by the United States and its allies and partners \citep{spacecom}.} The U.S. military's Space Surveillance Network detected over 1,500 trackable fragments within the first few days after the test \citep{spacecom} Modeling implied the presence of over 34,000 fragments too small to be tracked but large enough to cause catastrophic collisions. In early February 2022, monitoring services began to predict periodic ``squalls'' of up to thousands of new close approaches between C1408 fragments and active satellites per day, peaking in early April and recurring roughly every six months. These squalls were driven by the orbital mechanics of the debris fragments: as the fragments' orbits precessed along their inclinations, the debris overlapped the orbits of remote sensing satellites while moving in the opposite direction (i.e. with high risk of a catastrophic head-on collision). The large volume of close approach alerts produced greatly increased the environmental monitoring burden facing satellite operators, particularly those without automated maneuvering systems. The debris field is expected to persist for approximately two years, though each successive squall will be smaller as more fragments re-enter the Earth's atmosphere \citep{oltrogge2022iaa, oltrogge2022comparison}. The C1408 ASAT test is one of the largest and most destructive space debris events in history, and has sparked discussions about the need for international regulation and cooperation to prevent similar events in the future. \\

The relevant prior literature on space debris and orbit use can be broadly divided into two groups: engineering analyses and economic analyses. Engineering analyses have tended to focus on the dynamics of the debris fragments themselves and the risks they create for satellite operators. While this literature is rich, the most relevant analysis for this paper is \citet{lewis2010new}, which develops a network representation of the orbital environment using conjunction alerts (CAs) and the C1408 breakup. \citet{lewis2010new} uses this repreesntation with orbital trajectory propagators to calculate the likely evolution of the orbital environment, identify high-risk debris objects worth removing, and determine network statistics which can predict collision cascades and rank objects by removal priority. On the C1408 ASAT test itself, \citet{oltrogge2022iaa, oltrogge2022comparison, pardini2023short} analyze the evolution of the C1408 debris cloud both across the environment and with particular attention to inhabited space stations (i.e. the International Space Station and Tiangong) and large satellite constellations (i.e. Starlink). They find that the test substantially increased the debris flux in the orbital environment, though the fragment decay over time exceeded initial predictions. However, this literature has not yet considered how the C1408 ASAT test affected the relationships between different types of CAs or the effects of automated maneuvers on system-level resilience to cascades. \\

Economic analyses have focused on market failures, incentives leading to overproduction of debris and collision risk, and incentive-based policies to address the issue \citep{adilov2015economic, rouillon2020physico, 
grzelka2019managing, rao2020orbital, adilov2020economics}. These have largely been theoretical analyses, though \citet{rao2020orbital} and \citet{adilov2020economics} conduct calibrated simulations to assess the likely trajectory of the debris population into the future under different economic institutions and technology deployment scenarios. The general consensus is that orbital space is an example of an open-access commons, with the attendant rent dissipation and incentives to minimize effort spent on reducing environmental degradation. However, this literature has not conducted much empirical analysis of fragmentation events, nor has it analyzed the microstructure of the orbital environment and its behavior over time. \\

This paper uniquely bridges engineering and economic analyses of space debris, providing a detailed examination of the C1408 ASAT test's impact on the space traffic system and its systemic resilience. I build on \citet{lewis2010new} and existing economic literature on network games \citep{bramoulle2014strategic, galeotti2020targeting} to study space traffic network resilience to cascades using centrality measures. The exogeneity of the C1408 test, the difficulty of permanently altering a satellite's location, and a rich set of covariates and fixed effects enable causal identification of spillover effects of CAs with C1408 fragments, the influence of automated maneuvering capabilities on the network as a whole, and how these effects vary across the conjunction network. These contributions enhance our understanding of space traffic management, informing future policy and interventions aimed at ensuring the long-term sustainability of space activities. \\

The rest of the paper is organized as follows. Section \ref{sec:theory} outlines the details of the theoretical framework. It develops the idea of the conjunction network and the specific centrality measures used to assess system-level effects of the C1408 test. Section \ref{sec:empirics} develops the econometric framework for identifying the causal effects of the C1408 test on space traffic outcomes. Section \ref{sec:data} describes the data used to estimate the econometric models, as well as my criteria for identifying automated maneuvers from the CA data. Section \ref{results} presents econometric results and some discussion of their implications. I conclude in Section \ref{sec:conclusion}.

\section{Theoretical Framework}
\label{sec:theory}

In this section I present relevant background on fragmentation events and space traffic. These details shape the theoretical framework motivating my empirical analysis. There are two key ideas behind the framework. 
First, close approaches (conjunctions) between orbiting objects reflect their proximity and induce a time-varying network connecting them. Since distance between objects in physical space has a difficult-to-summarize non-Euclidean structure due to orbital mechanics, it is analytically convenient to work with the topology of the conjunction network instead. 
Second, while satellite operators tend to exert effort to avoid collisions (e.g. by coordinating maneuvers with their conjunction partners when possible), there is always the potential of a collision with their conjunction partners due to technical or human error. These collisions can disrupt the conjunction network, propagating risk or causing structural changes to the network.

\subsection{Fragmentation events and space traffic}

When satellite fragmentation events occur, such as ASAT tests, the resulting fragment cloud disperses in an ever-expanding manner. As it stretches across a larger volume, the spatial density of debris diminishes. However, trackable fragments are not the sole concern---lethal non-trackable (LNT) fragments also lurk within these debris fields, posing additional threats to operational satellites.\footnote{LNTs are just what the name suggests: fragments too small to be tracked (i.e. at the time of this writing, smaller than 10 cm in diameter), but large enough to cause lethal damage to a payload.} Consequently, satellite operators must invest extra effort in screening for potential LNT risks when evaluating conjunction alerts involving C1408 fragments, as they bring about additional operational costs. \\ 
 

Conjunction alerts (CAs) play a crucial role in satellite operations, providing operators with valuable information about potential close approaches and collisions between orbiting objects. Central to these alerts is the collision probability ($Pc$), an essential metric estimating the chances of a collision between two objects. To generate CAs, propagators rely on position data from sensor networks to predict orbital trajectories. These trajectories tend to have significant uncertainty, particularly as the time since the position data measurement increases, due to orbital perturbations caused by influences such as lunar gravity, nonuniformities in the Earth's surface, and atmospheric drag. Uncertainties may also arise from sensor limitations and limitations in trajectory propagation algorithms. These uncertainties necessitate constant updates to trajectory predictions. \\

When confronted with numerous CAs, operators face increased costs in terms of time, computational resource requirements, longer propagator runtimes, reduced flight safety notification bandwidth, and lower overall spacecraft safety. These challenges scale superlinearly with the number of satellites operating in a given volume.\footnote{A rough-but-empirically-useful rule of thumb is that CAs scale quadratically with the number of objects in a thin spherical shell. See \citet{liou2009sensitivity, lewis2017sensitivity} for more on how engineers tend to model the orbital environment.} To mitigate these challenges, large satellite fleets often use automated maneuvering systems to reduce the burden on human operators. These systems establish rules for determining when a CA warrants a maneuver, streamlining the decision-making process and allowing fleets to scale beyond what a relatively small number of humans can manage.

\subsection{The conjunction network}

I follow \citet{lewis2010new} in developing my analysis of the orbital environment from the network induced by CAs, i.e. the ``conjunction network''. \citet{lewis2010new} finds that the conjunction network can provide insight into the importance and vulnerability of individual satellites in the orbital environment. The conjunction network approach can be particularly useful in identifying key objects to target for removal or mitigation. For example, the analysis of the conjunction network can reveal objects with high betweenness centrality, which are critical to the flow of conjunction events between satellites and thus the risk of collision cascades. Furthermore, \citet{lewis2010new} find evidence of disassortative mixing in the conjunction network, suggesting that a targeted approach to debris removal or satellite maneuvering would be more effective than random removal or ad hoc maneuvering, as highly connected objects are the most vulnerable points in the network. \\

There are $N(t)$ objects in an orbital volume at time $t$, connected by the $N(t) \times N(t)$ conjunction matrix $C(t)$ with typical element $c_{ij}(t) \geq 0$ measuring the frequency of close encounters between objects $i$ and $j$.\footnote{``Close'' here means ``less than a specified distance between object centers of mass''. As the specified distance increases all elements of $C(t)$ become strictly positive. As I describe later, my data is generated using a distance of 5 km between object centers of mass. The objects I study inhabit the region between roughly 200 and 2000 km above mean sea level, so the specified distance is relatively small in comparison.} The conjunction matrix provides a way to summarize the topology of the network of close approaches (conjunctions) between orbiting objects. Each entry $c_{ij}(t)$ is a proxy for the probability of a close approach between objects $i$ and $j$ at time $t$. \\

The connection degree distribution of the conjunction matrix also provides a convenient way to measure the proximity between objects. Two objects $i$ and $j$ are first-degree neighbors if $c_{ij} > 0$. Objects $i$ and $k$ are second-degree neighbors if the shortest path between them is through $j$ and $c_{ij} c_{jk} > 0$, and so on for higher-degree neighborhoods. Two objects with lower-degree connections to each other are closer over the network's topology than two objects with higher-degree connections. The topology of the conjunction network can change over time as new close approaches occur, existing links are reinforced or weakened, and new links are formed or broken. Because the conjunction matrix captures the patterns of space traffic and likely volumetric expansion paths following fragmentation events, it can be used to study how traffic disruptions propagate and systemic resilience evolves over space and time. To study how the disruption caused by CAs with C1408 fragments varies over the pre-ASAT test network structure, I calculate each object's network distance from C1408 in the month prior to its destruction---I refer to these distances as the ``C1408 connection neighborhoods.'' I describe the construction of the C1408 connection neighborhoods in more detail in Section \ref{sec:data}. \\


``Spillover effects'' in the context of the conjunction network refer to self-induced changes in the magnitude or structure of the conjunction matrix $C(t)$, i.e. changes in the frequency of close approaches between two objects $i$ and $j$ that arise due to changes in the frequency of close approaches involving either object $i$ or object $j$ themselves. For example, changes in the frequency of conjunctions between (individual or types of) objects $i$ and $j$ due to conjunctions between objects $j$ and $k$ are self-induced changes, i.e. derivatives of the form $\pdv{c_{ij}}{c_{jk}}$. Spillover effects can propagate through the network, amplifying or dampening the risk of collision cascades. I focus on estimating spillover effects on object $i$'s Other CAs caused by C1408 fragment CAs, i.e. derivatives of the form $\pdv{c_{i,Other}}{c_{j,C1408}}$.

\subsection{Centrality}

Centrality is a measure of the importance of nodes in a network. It captures the extent to which nodes connect to other nodes and the importance of those other nodes. In the context of the orbital environment and conjunction network, centrality can be used to identify important and vulnerable objects for removal or mitigation \citep{lewis2010new} Centrality is often associated with viral phenomena and cascades because highly central nodes can spread information or influence more efficiently throughout the network \citep{keeling2005networks, danon2011networks, bedson2021review}. In network games with strategic interactions or externalities, centrality measures can indicate the relative influence of different nodes and how changes in incentives or actions of one node can affect the actions of others in the network \citep{bramoulle2014strategic, galeotti2020targeting}. I focus on two specific measures: betweenness and eigenvector centrality. The former has been found to be relevant to collision cascades in orbit, while the latter has been shown to be important to determining optimal targeted interventions in games played over networks.

\paragraph{Betweenness centrality:} The betweenness centrality of the conjunction matrix $C(t)$ measures the degree to which an individual object is important for maintaining network connectivity. It can be defined as the fraction of shortest paths between any two objects in the network that pass through a given object. Formally, the betweenness centrality $b_i$ of object $i$ at time $t$ is given by:

\begin{equation*}
b_i (t) = \sum_{j \neq i \neq k} \frac{\sigma_{jk}(i, t)}{\sigma_{jk}(t)}
\end{equation*}

where $\sigma_{jk}(t)$ is the total number of shortest paths through $C(t)$ between objects $j$ and $k$ at time $t$, and $\sigma_{jk}(i, t)$ is the number of shortest paths between $j$ and $k$ that pass through object $i$ at time $t$. \\

Betweenness centrality is informative about the risk of collision cascades due to collisions between satellites that are not highly-connected but are instead located at important positions in the network. Increases in average betweenness centrality suggest that either new objects/connections have been added to the network (e.g. due to detection of existing objects or entry of new objects), or that there are more objects connecting distinct sub-networks. In the latter case the increase in betweenness centrality indicates that a collision cascade is more likely to spread across otherwise-distant object groups. My choice of the betweenness centrality measure is motivated by \citet{lewis2010new}'s analysis of conjunction networks. \citet{lewis2010new} conduct Monte Carlo simulations of the evolution of the debris environment to study which statistics of the conjunction network predicted overall collision outcomes. They find that betweenness centrality can be used to measure the risk of a collision cascade in the space debris network, and that removing objects with high betweenness centrality is an efficient way to prevent cascades.

\paragraph{Eigenvector centrality:} The eigenvector centrality of $C(t)$ measures the degree of connectedness between highly-connected nodes. Formally, the eigenvector centrality $\lambda_i(t)$ of an object $i$ at time $t$ is the $i$th entry of $\lambda(t)$, which satisfies
\begin{equation}
C(t) \mathbf{v}(t) = \lambda(t) \mathbf{v}(t)
\label{eqn:eigen-centrality}
\end{equation}

for eigenvectors $\mathbf{v}(t)$. Equation \ref{eqn:eigen-centrality} shows that the eigenvector centrality of each object is proportional to the sum of the eigenvector centralities of its neighbors, weighted by the strength of their connections. Objects with high eigenvector centrality are highly connected to other highly central objects. \\

Damaging collisions or fragmentation events affecting these objects are likely to affect many other objects across the network. Further, under the ID orbit model, these objects may be more likely to experience collisions in the first place. Increases in the eigenvector centrality indicate periods with higher risk of collision cascades. Protecting, avoiding, or deorbiting objects with high eigenvector centrality may be an effective way to prevent or mitigate collision cascades. My choice of eigenvector centrality is motivated by the economic literature on network games. \citet{bramoulle2014strategic} show that eigenvalues of the network's adjacency matrix capture the degree to which the network amplifies agents' actions. \citet{galeotti2020targeting} show that when networks mediate strategic spillovers and externalities among players in a game, optimal targeted interventions to alter players' investments will be determined by their eigenvector centralities. In the space traffic context, the eigenvector centralities of the conjunction network could be used to determine which satellites are most critical for the stability of the network, and thus which ones should be targeted for collision avoidance measures to prevent or mitigate collision cascades. Targeting satellites by eigenvector centrality can magnify the impacts of interventions through the spillover effects of their connections with other satellites and debris in the network.

\section{Empirical Framework}
\label{sec:empirics}

I index payloads by $i$ and days by $t$. In all models $CA^{k}_{it}$ refers to CAs with objects of type $k$ (C1408 or Other), $\lambda_{it}$ and $b_{it}$ refer to eigenvector and betweenness centralities, and $M_{it}$ refers to the estimated number of automated maneuvers conducted. $X_{it}$ refers to a set of potentially payload-day level covariates (e.g. log mean estimated collision probability, $\log(E[Pc_{it}])$). $\gamma_{it}$ refers to fixed effects across payloads, time, or some groupings thereof (e.g. pre/post C1408 ASAT test, C1408 connection neighborhood, satellite ID). Where covariates or fixed effects only vary across a single dimension for a family of models, only that subscript is shown. The coefficient of interest is always identified as $\beta$. \\

To explore how CAs with C1408 fragments affected CAs with non-C1408 objects, I estimate ``spillover models'' of the form in equation \ref{eqn:crowdout-model-family}:
\begin{equation}
    CA^{Other}_{it} = \beta CA^{C1408}_{it} + \delta X_{it} + \gamma_{it} + \epsilon_{it}. \label{eqn:crowdout-model-family}
\end{equation}

I first estimate the spillover models for the full sample, and then separately by C1408 connection neighborhood. My full specification for the spillover models includes the log of the mean estimated collision probability to control for changes in object state and local volume characteristics, and lagged conjunctions with C1408 fragments and Other fragments to control for serial correlations. I also include fixed effects for pre/post the ASAT test, the connection neighborhood (when estimated over the full sample), and satellite ID (which controls for the satellite's average locational characteristics). To explore how the effects of C1408 CAs vary by C1408 connection neighborhood, I also estimate the full specification with C1408 CAs interacted with connection neighborhood. \\

To explore how automated maneuvers affect network resilience, I separately estimate ``centrality models'' of the form in equations \ref{eqn:resilience-model-family-between} and \ref{eqn:resilience-model-family-eigen}:
\begin{align}
    b_{it} &= \beta M_{it} + \delta X_{it} + \gamma_{it} + \epsilon_{it}, \label{eqn:resilience-model-family-between} \\
    \lambda_{it} &= \beta M_{it} + \delta X_{it} + \gamma_{it} + \epsilon_{it},\label{eqn:resilience-model-family-eigen}
\end{align}

where in some specifications $X_{it}$ includes $CA^{C1408}_{it}$ and $CA^{Other}_{it}$. I first estimate the centrality models for the full sample, and then separately by C1408 connection neighborhood. My full specification for the centrality models includes the same covariates and fixed effects as the spillover models, plus Other CAs. To make the $\beta$ coefficients easier to interpret, I standardize both centralities to have mean 0 and standard deviation 1. As with the spillover models, I also estimate the full specification with C1408 CAs interacted with connection neighborhood. \\


\section{Data}
\label{sec:data}
\begin{figure}[htbp] 
	\centering
	\includegraphics[width=\textwidth]{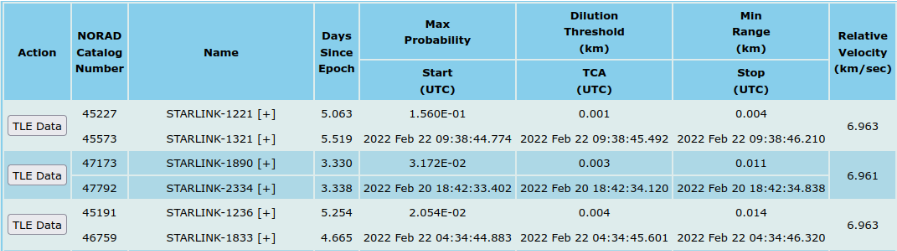} 
	\caption{Example of CAs issued by the SOCRATES system.
	}
	\label{fig:cdm-example}
\end{figure}

The main source of data for this analysis is predictions of close approaches, known as ``conjunction alerts'' (``CAs''), between objects in low-Earth orbit. I obtain conjunction alert data from the SOCRATES (Satellite Orbital Conjunction Reports Assessing Threatening Encounters in Space) service provided by CelesTrak \citep{socrates}. Three times each day, SOCRATES collects data on orbital object positions, simulates them forward in time for the next week, and reports all predicted approaches within 5 km at the time of closest approach between two objects. The object position data are collected from publicly available files provided by the US Government on the Space-Track.org system, which is maintained by the 18th Space Defense Squadron \citep{spacetrack}, and simulated forward using the SGP4 numerical propagator (a commonly-used tool for this type of task). Each CA includes the names and catalog ID numbers of the objects involved and prediction details such as the estimated probability of a collision if no action is taken, the time of closest approach, and the distance at the time of closest approach.\footnote{The format of the CA is called a ``Conjunction Data Message'', or CDM.} Figure \ref{fig:cdm-example} shows sample CAs reported by the SOCRATES system's web interface. The dataset does not include details on the objects' physical locations or trajectories at the time of the prediction.\footnote{I am currently working on collecting and linking these data from the historical Space-Track.org data.} The NORAD Catalog Number uniquely identifies each tracked object in orbit, so I refer to it in tables as ``Satellite ID''. The SOCRATES and Space-Track.org data are available for free online. \\ 

The CAs are recorded at the conjunction pair-level, i.e. each unique CA corresponds to a specific pair of objects predicted to have a close approach at a particular time. The primary object is always a satellite, while the secondary object may be another satellite or a debris object. I focus on the period between 15 May 2021 and 15 November 2022, six months before and one year after the C1408 event (15 November 2021). Over this period there were over 39 million unique CAs (39,765,335) recorded in the SOCRATES system. These CAs represent the entire population of tracked objects in orbit over this time.\footnote{Existing sensors can detect objects as small as roughly 10 cm in diameter. While there are statistical characterizations of the populations of smaller objects, they are not represented in CA datasets.} I aggregate conjunctions to the daily-primary object level, giving a sample of just over 3 million daily records for primary objects (3,119,081). \\

\begin{figure}[htbp] 
	\centering
	\includegraphics[height=0.8\textheight]{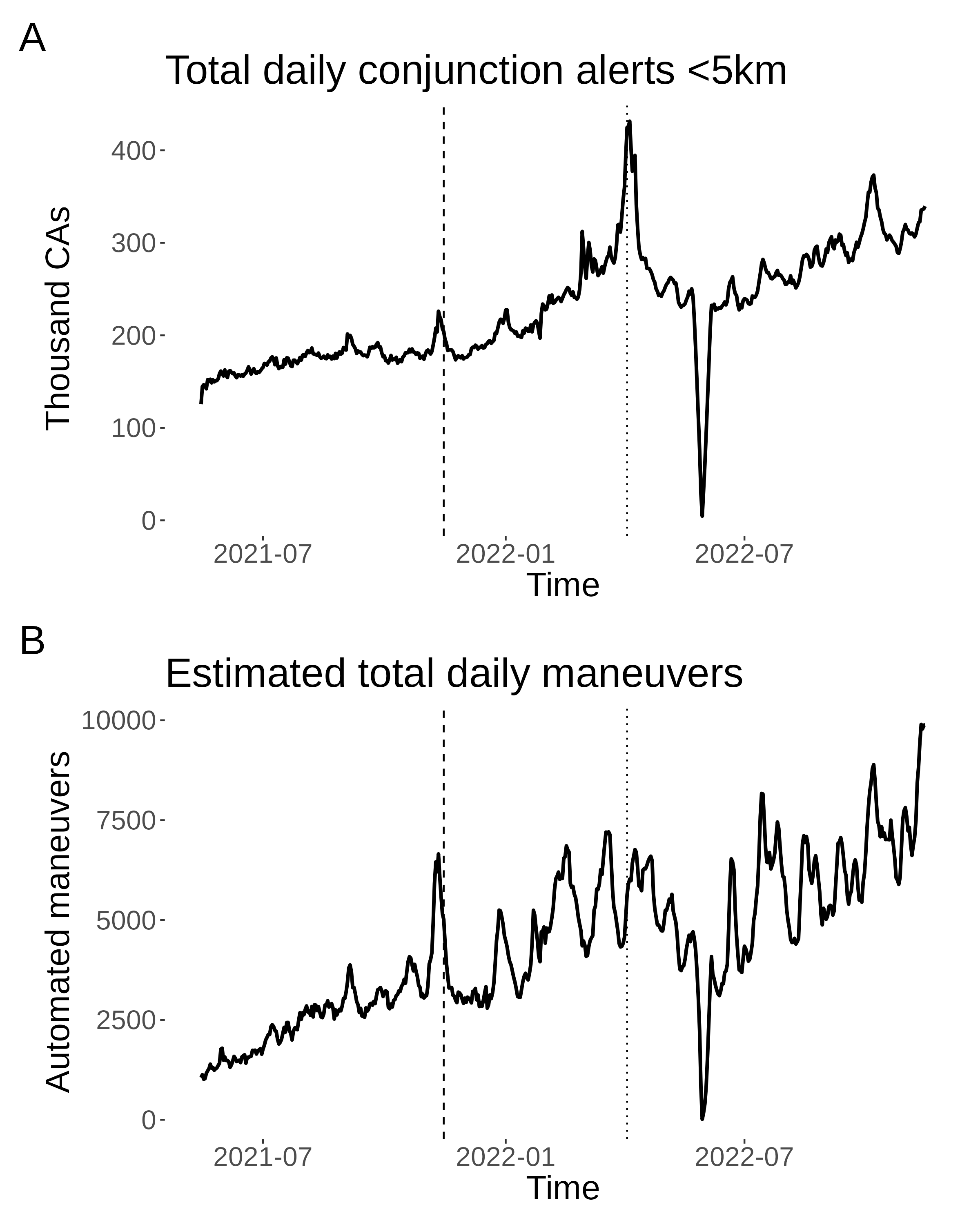} 
	\caption{Daily total CA and estimated automated maneuver counts, the latter based on publicly-announced maneuver criteria. The dashed line shows the date of the C1408 ASAT test, and the dotted line shows the date when the first debris squall peaked. The large dip in late May/early June is due to missing data.
	}
	\label{fig:sample-CAs-maneuvers-plot}
\end{figure}

I attribute maneuvers by the two operators who have publicly announced their use of automated maneuvering systems---Planet and SpaceX---based on their publicly-announced maneuver thresholds. Both operators have stated that their systems will conduct an automated maneuver if the estimated collision probability (referred to as ``$Pc$'' in the space debris modeling literature) exceeds a threshold---Planet will maneuver if the probability exceeds $10^{-4}$ while SpaceX will maneuver if the probability exceeds $10^{-5}$ \citep{planetDetails, spacexDetails}.\footnote{The Inter-Agency Space Debris Coordination Committee (IADC) is an inter-governmental forum which aims to coordinate space debris management and mitigation \citep{iadc}. While it does not have any enforcement or rules-issuing authority, its latest guidelines recommend maneuvers if the collision probability exceeds $10^{-4}$. Both operators have stated that they primarily use automated maneuvering systems but have also expressed willingness to disable their automated systems for specific conjunctions if their conjunction partner would prefer that. These discussions can sometimes be fraught, as operators may use different data sources to compute collision probabilities \citep{starlinkOneweb}.} I do not observe maneuvers by operators using non-automated systems, which are often conducted based on ad hoc criteria. Table \ref{tab:agg-sum-stats} shows aggregate summary statistics for CAs and maneuvers over my sample, and figure \ref{fig:sample-CAs-maneuvers-plot} plots the daily total CA and estimated automated maneuver counts. There is a large dip in CAs in late May/early June due to a glitch in the SOCRATES system which led to missing data. Since the SOCRATES system is not the primary source of CAs for operators---there are other data providers who offer (paid) higher quality-of-service, while SOCRATES is free but offers lower quality-of-service---this data outage did not affect operator behavior. \\

\begin{table}
\caption{Aggregate summary statistics}
\centering
\begin{tabular}[t]{rrrrr}
\toprule
Sample size & Maneuvers (Planet) & Maneuvers (Starlink) & C1408 CAs & Other CAs\\
\midrule
2,730,865 & 10,439 & 2,113,161 & 9,499,864 & 103,975,720\\
\bottomrule
\end{tabular}
\label{tab:agg-sum-stats}
\end{table}

My analysis focuses on objects which were in a connected component with C1408 in the month prior to its destruction (15-10-2022--14-11-2022). This includes objects which conjuncted with C1408 during that period (``C1408's first-degree neighbors''); objects which conjuncted with those objects (``C1408's second-degree neighbors''); and so on. C1408 had neighbors up to 5 network links away, i.e. up to fifth-degree neighbors. I refer to these groups as ``neighborhood groups''. Since the fifth-degree neighborhood consists of just 6 object-day observations on two objects, I drop this group. I also drop the group that isn't connected to C1408 at all, which contains 391 objects with 389,213 object-day observations. My final dataset is a panel with 6,607 primary satellite objects for analysis over 549 days. Given that some objects deorbited during this period, my sample contains a little over 2.7 million object-day observations. \\

This sample represents the minimal set of objects whose space traffic patterns could plausibly have been disrupted by C1408; while some of the objects that were initially disconnected from C1408 could have been affected, this sample lets me study how the effects of C1408's destruction propagated along the network structure that existed at the time of its destruction. No objects were observed to experience a collision with any tracked C1408 fragments.\footnote{Some objects may have experienced collisions with untrackable fragments and suffered operational issues; if this happened the specifics have not been publicly disclosed at the time of this writing.} Table \ref{tab:group-sum-stats} shows summary statistics for CAs, maneuvers, and centrality measures by C1408 neighbor grouping. Note that since the centrality measures are between 0 and 1, the standard deviations in parentheses reflect greater variability above than below the mean. \\

\begin{table}[ht]
\centering
\caption{Summary statistics by C1408 neighborhood group}
\begin{adjustbox}{max width=1.15\textwidth}
\hspace*{-2cm}
\begin{tabular}{@{}cccccccccSSS@{}}
\toprule
\multirow{2}{*}{\shortstack{C1408 \\ neighborhood}} & \multirow{2}{*}{\shortstack{Sample \\ size}} & \multirow{2}{*}{\shortstack{Satellites}} & \multicolumn{2}{c}{Maneuvers} & \multicolumn{2}{c}{CAs} & \multicolumn{2}{c}{Mean (SD) centralities}  \\
\cmidrule(lr){4-5} \cmidrule(lr){6-7} \cmidrule(lr){8-9} 
& & & Planet & Starlink & C1408 & Other & Eigenvector & Betweenness \\
\midrule
\multirow{2}{*}{1} & \multirow{2}{*}{53,576} & \multirow{2}{*}{103} & 916 & 2,476 & 879,568 & 1,088,254 & \shortstack{0.0116 \\ (\num{0.0359})} & \shortstack{0.0004 \\ (\num{0.0008})} \\
& & & & & & & & \\
\multirow{2}{*}{2} & \multirow{2}{*}{1,541,232} & \multirow{2}{*}{2,980} & 8,513 & 1,864,681 & 7,846,341 & 58,150,972 & \shortstack{0.0632 \\ (\num{0.0943})} & \shortstack{0.0005 \\ (\num{0.0010})} \\
& & & & & & & & \\
\multirow{2}{*}{3} & \multirow{2}{*}{1,121,313} & \multirow{2}{*}{2,798} & 1,006 & 245,836 & 772,120 & 44,424,221 & \shortstack{0.0111 \\ (\num{0.0447})} & \shortstack{0.0006 \\ (\num{0.0013})} \\
& & & & & & & & \\
\multirow{2}{*}{4} & \multirow{2}{*}{12,896} & \multirow{2}{*}{196} & 0 & 39 & 76 & 264,227 & \shortstack{0.0004 \\ (\num{0.0124})} & \shortstack{0.0001 \\ (\num{0.0004})} \\
& & & & & & & & \\
\bottomrule
\end{tabular}
\end{adjustbox}
\label{tab:group-sum-stats}
\end{table}

In table \ref{tab:group-sum-stats}, it is clear that C1408's second- and third-degree neighborhoods contain the most objects. This is partly a reflection of C1408's position in the network, and partly a feature of the network structure overall. Most of the Starlink and Planet satellites are located in the second-degree neighborhood. Given that these satellites often experience CAs within the constellation (e.g. Starlink satellites with other Starlink satellites), they will naturally be very highly connected to each other within the constellation. The degree of connection depends on their constellation architecture: Starlink satellites must cover the globe at every instant to provide telecommunications service, while Planet satellites only need to cover the entire globe at least once per day. The Starlink system is also considerably larger than the Planet system, having just over 3,000 satellites at the time of this writing (and growing rapidly), whereas Planet has fewer than 400 in total.




\section{Results}
\label{sec:results}

\begin{figure}[htbp] 
	\centering
	\includegraphics[width=\textwidth]{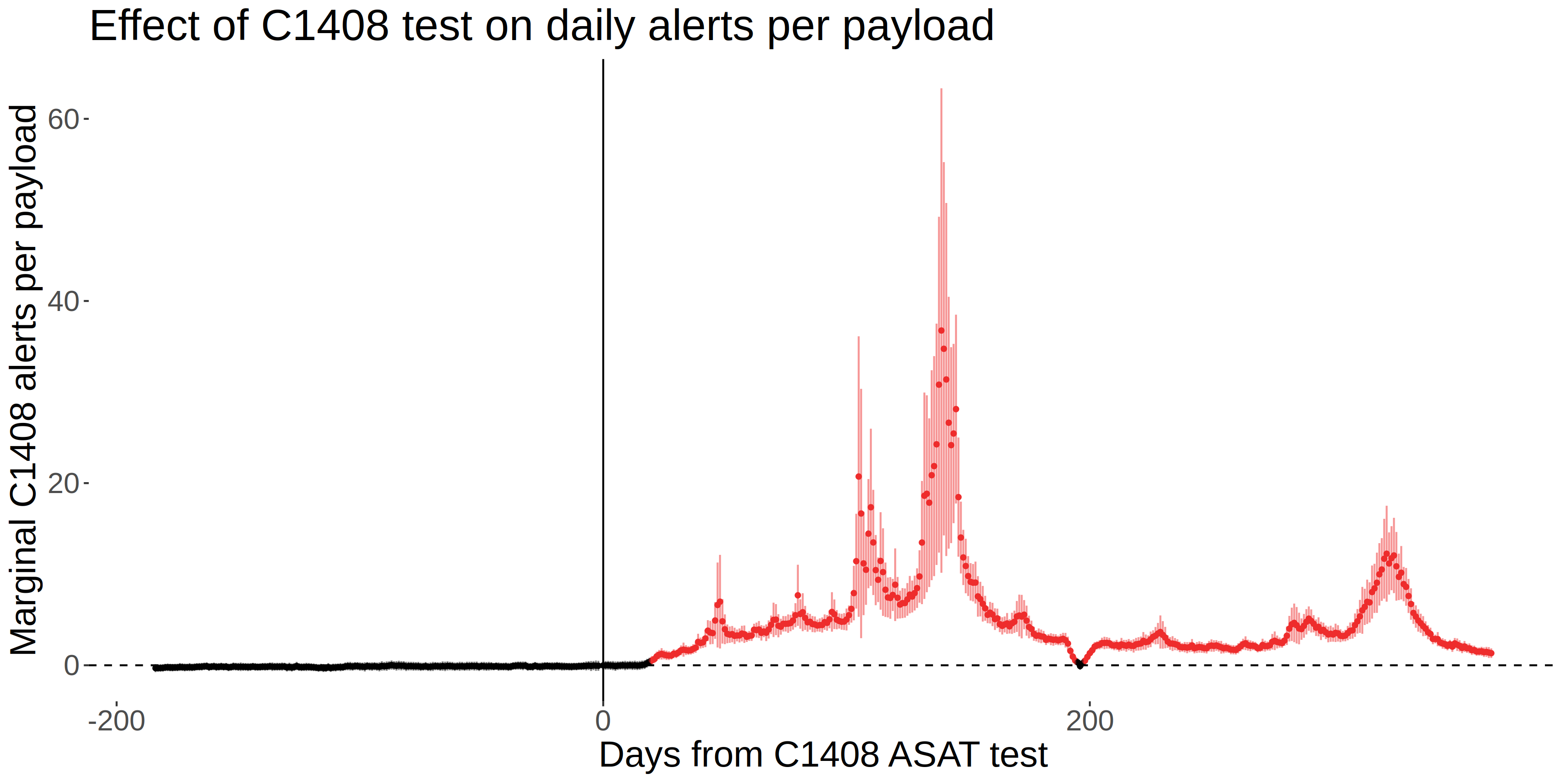} 
	\caption{Daily marginal CAs per payload with C1408 fragments following the ASAT test, controlling for payload-specific C1408 exposure. The decline in CAs around day 200 is due to an exogenous glitch in the SOCRATES system which caused missing data.
	}
	\label{fig:c1408-CAs-eventstudy-full}
\end{figure}

I use equation \ref{eqn:c1408-CAs-eventstudy-full} below to calculate the daily average marginal CA count per payload relative to the day before the ASAT test over the full sample (15-05-2021--15-11-2022). Figure \ref{fig:c1408-CAs-eventstudy-full} shows daily marginal CAs per payload with C1408 fragments following the ASAT test, i.e. the $\beta_j$ coefficients from equation \ref{eqn:c1408-CAs-eventstudy-full}. The $\gamma_i$ are primary payload-fixed effects to control for payload-specific exposure to C1408 fragments. The debris squalls around 150 and 350 days after the ASAT test are visible as increases in the marginal C1408 conjunctions. While these regressions have the structure of an event study model, I use them to generate descriptive facts about the size of the debris squalls and the validity of network distance in this context.\footnote{The estimates are also causal, but ``it's causal because the fragments didn't exist till C1408 got blown up'' is not a very exciting identification argument for a fairly obvious statement.}

\begin{equation}
    CA^{C1408}_{it} = \sum_j \beta_j ASAT_{t-j} + \gamma_i + \epsilon_{it} \label{eqn:c1408-CAs-eventstudy-full}
\end{equation}

\begin{figure}[htbp] 
	\centering
	\includegraphics[width=\textwidth]{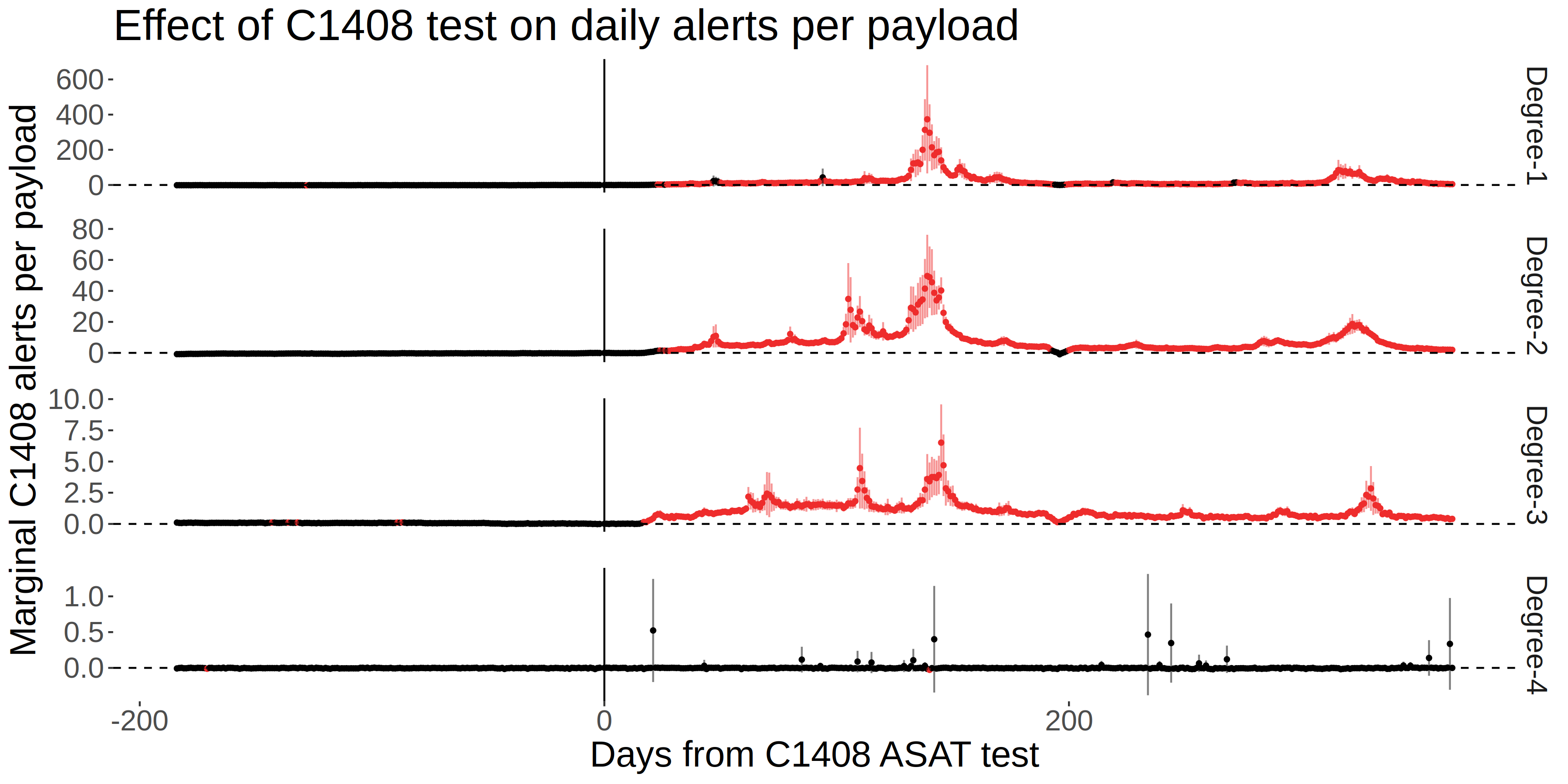} 
	\caption{Daily marginal CAs per payload with C1408 fragments following the ASAT test, controlling for payload-specific C1408 exposure, split by C1408 connection neighborhood.
	}
	\label{fig:c1408-CAs-eventstudy-split}
\end{figure}

To explore how the CA burden varies by pre-ASAT test network distance from C1408, I estimate equation \ref{eqn:c1408-CAs-eventstudy-full} separately for each C1408 connection neighborhood. Figure \ref{fig:c1408-CAs-eventstudy-split} shows these effects. As before, the debris squalls around 150 and 350 days after the ASAT test are visible as increases in the marginal C1408 conjunctions. While the patterns are similar, the scale of C1408 fragment exposure varies substantially by connection neighborhood. The first neighborhood experiences the most severe effects---around 400 marginal CAs per payload at the first squall peak. These effects decay sharply over network distance, with the fourth neighborhood only occasionally being exposed to a C1408 fragment. I interpret this as validation of the connection neighborhood approach: network topology appears to reflect a physically-meaningful distance metric. \\

With descriptive evidence to contextualize the size of the debris squalls and validate network distance as a meaningful dimension of effect heterogeneity, I turn to the spillover and centrality models. Table \ref{tab:crowdout-models-results} shows results from estimating a series of spillover models described in equation \ref{eqn:crowdout-model-family}. The results are consistent with operators maneuvering when C1408 fragments get close, i.e. C1408 is disrupting traffic but operators have enough space to avoid each other.\footnote{Since the model does not incorporate maneuvers by all operators (i.e. non-automated maneuvers are not included) this is not the only possible explanation. It is also possible that sensor networks were overloaded by the volume of objects being tracked and focused on predicting C1408 CAs at the expense of not predicting Other CAs. While either possibility is consistent with Other CAs being crowded out rather than crowded in, they have different implications for the capacity of the space traffic system. The former would indicate the presence of additional physical space for safety during disruptions, while the latter would indicate a lack of systemic capacity to monitor large volumes of interactions. Though this would be a concerning possibility, neither operators of the SOCRATES system nor the operators of the Space-Track.org system have reported any C1408-related CA triage.} The magnitude of the spillover effect decays monotonically with network distance from C1408 over the first three C1408 connection neighborhoods, with the effect at the fourth neighborhood being both large and statistically insignificant. To place these results in context, the April squall peak involved over 400 marginal CAs per payload in the first C1408 connection neighborhood in a single day. Results from column (5) of table \label{tab:crowdout-models-results} suggest that on average each payload in that neighborhood faced 2 fewer Other CAs as a result. \\

\begin{table}[htbp]
\label{tab:crowdout-models-results}
\caption{Spillover models: Other CAs as outcome}
\centering
\begin{adjustbox}{max width=1.15\textwidth}
\begin{tabular}[t]{lccccc}
\toprule
  & (1) & (2) & (3) & (4) & (5)\\
\midrule
$CA^{C1408}_{it}$ & \num{-0.034}*** & \num{-0.009}*** & \num{-0.006}*** & \num{-0.002}* & \\
 & (\num{0.005}) & (\num{0.002}) & (\num{0.001}) & (\num{0.001}) & \\
$CA^{C1408}_{it-1}$ &  & \num{-0.008}*** & \num{-0.004}*** & \num{0.001} & \num{0.001}\\
 &  & (\num{0.001}) & (\num{0.001}) & (\num{0.001}) & (\num{0.001})\\
$CA^{Other}_{it-1}$ &  & \num{0.592}*** & \num{0.569}*** & \num{0.412}*** & \num{0.412}***\\
 &  & (\num{0.011}) & (\num{0.010}) & (\num{0.013}) & (\num{0.014})\\
$\log(E[Pc_{it}])$ &  &  & \num{4.348}*** & \num{4.598}*** & \num{4.598}***\\
 &  &  & (\num{0.081}) & (\num{0.069}) & (\num{0.051})\\
$CA^{C1408}_{it} \times$ C1408--1 &  &  &  &  & \num{-0.005}***\\
 &  &  &  &  & \vphantom{1} (\num{0.001})\\
$CA^{C1408}_{it} \times$ C1408--2 &  &  &  &  & \num{-0.002}**\\
 &  &  &  &  & (\num{0.001})\\
$CA^{C1408}_{it} \times$ C1408--3 &  &  &  &  & \num{0.005}\\
 &  &  &  &  & (\num{0.004})\\
$CA^{C1408}_{it} \times$ C1408--4 &  &  &  &  & \num{-0.808}\\
 &  &  &  &  & (\num{0.455})\\
\midrule
Num.Obs. & \num{2730872} & \num{2514877} & \num{2514877} & \num{2514877} & \num{2514877}\\
FE: pre/post ASAT & X & X & X & X & X\\
FE: Satellite ID &  &  &  & X & X\\
FE: C1408 neighborhood & X & X & X & X & \\
\bottomrule
\multicolumn{6}{l}{\rule{0pt}{1em}* p $<$ 0.05, ** p $<$ 0.01, *** p $<$ 0.001.}\\
\multicolumn{6}{l}{\rule{0pt}{1em}C1408--$n$ refers to interactions with the $n^{th}$ C1408 connection neighborhood.}\\
\multicolumn{6}{l}{\rule{0pt}{1em}Standard errors in parentheses. All standard errors are Driscoll-Kraay with $L=4$.}\\
\end{tabular}
\end{adjustbox}
\end{table}

Next, I turn to the centrality models. Table \ref{tab:centrality-models-results-between} shows results from estimating a series of centrality models described in equation \ref{eqn:resilience-model-family-between}. The results indicate that automated maneuvers reduce betweenness centrality by small but statistically significant amounts. These estimates are consistent with automated maneuvers limiting the potential for collision cascade---i.e. they seem to increase the resilience of the space traffic system. While the estimates are small in magnitude, they are likely economically significant given the high volume of automated maneuvers conducted by Planet and Starlink---as high as nearly 2 million automated maneuvers over the sample period in the second connection neighborhood (see table \ref{tab:group-sum-stats}). While centrality estimates cannot be summed, small improvements in traffic system resilience across a large number of actions are still notable. As with the spillover effects, column (5) shows that the effects decay monotonically with network distance from C1408 until the fourth connection neighborhood, which is economically large, positive, and statistically insignificant. \\

\begin{table}[htbp]
\label{tab:centrality-models-results-between}
\caption{Centrality models: Betweenness centrality as outcome}
\centering
\begin{adjustbox}{max width=1.15\textwidth}
\begin{tabular}[t]{lccccc}
\toprule
  & (1) & (2) & (3) & (4) & (5)\\
\midrule
$M_{it}$ & \num{-0.016}*** & \num{-0.054}*** & \num{-0.057}*** & \num{-0.035}*** & \\
 & (\num{0.001}) & (\num{0.002}) & (\num{0.002}) & (\num{0.002}) & \\
$CA^{C1408}_{it}$ &  & \num{0.002}*** & \num{0.002}*** & \num{0.002}*** & \num{0.002}***\\
 &  & (\num{0.000}) & (\num{0.000}) & (\num{0.000}) & \vphantom{3} (\num{0.000})\\
$CA^{Other}_{it}$ &  & \num{0.005}*** & \num{0.005}*** & \num{0.005}*** & \num{0.005}***\\
 &  & (\num{0.000}) & (\num{0.000}) & (\num{0.000}) & \vphantom{2} (\num{0.000})\\
$CA^{C1408}_{it-1}$ &  & \num{0.000}** & \num{0.000}** & \num{0.001}*** & \num{0.001}***\\
 &  & (\num{0.000}) & (\num{0.000}) & (\num{0.000}) & \vphantom{1} (\num{0.000})\\
$CA^{Other}_{it-1}$ &  & \num{0.001}*** & \num{0.001}*** & \num{0.001}*** & \num{0.001}***\\
 &  & (\num{0.000}) & (\num{0.000}) & (\num{0.000}) & (\num{0.000})\\
$\log(E[Pc_{it}])$ &  &  & \num{0.010}*** & \num{0.031}*** & \num{0.031}***\\
 &  &  & (\num{0.002}) & (\num{0.002}) & (\num{0.002})\\
$M_{it} \times$ C1408--1 &  &  &  &  & \num{-0.047}***\\
 &  &  &  &  & (\num{0.011})\\
$M_{it} \times$ C1408--2 &  &  &  &  & \num{-0.036}***\\
 &  &  &  &  & \vphantom{1} (\num{0.002})\\
$M_{it} \times$ C1408--3 &  &  &  &  & \num{-0.026}***\\
 &  &  &  &  & (\num{0.002})\\
$M_{it} \times$ C1408--4 &  &  &  &  & \num{0.242}\\
 &  &  &  &  & (\num{0.282})\\
\midrule
Num.Obs. & \num{2729030} & \num{2513281} & \num{2513281} & \num{2513281} & \num{2513281}\\
FE: pre/post ASAT & X & X & X & X & X\\
FE: Satellite ID &  &  &  & X & X\\
FE: C1408 neighborhood & X & X & X & X & \\
\bottomrule
\multicolumn{6}{l}{\rule{0pt}{1em}* p $<$ 0.05, ** p $<$ 0.01, *** p $<$ 0.001.}\\
\multicolumn{6}{l}{\rule{0pt}{1em}Standard errors in parentheses. All standard errors are Driscoll-Kraay with $L=4$.}\\
\end{tabular}
\end{adjustbox}
\end{table}

Finally, table \ref{tab:centrality-models-results-eigen} shows results from estimating a series of centrality models described in equation \ref{eqn:resilience-model-family-eigen}. The results indicate that automated maneuvers increase eigenvector centrality by small but statistically significant amounts. The implications of these estimates is less clear than those for betweenness centrality---while the argument for their likely economic significance is similar (small changes occurring many times), whether this is good or bad for network resilience depends on the nature of the game being played. \citet{galeotti2020targeting} find that optimal targeted interventions in network games are determined by eigenvector centrality. In games of strategic substitutes (my action makes you less likely to take an action), optimal interventions target agents with lower eigenvector centrality as their local network structure makes their actions less likely to crowd out others' actions. In games of strategic complements (my action makes you more likely to take an action), optimal interventions target agents with higher eigenvector centrality as their local network structure makes their actions more likely to crowd in others' actions. Space traffic management can be exhibit both strategic substitutability or strategic complementarity depending on the situation. When there is plenty of room to maneuver, maneuvers are strategic substitutes because CAs and collisions are reciprocal: if I maneuver out of the way, you no longer need to do so. Similarly, debris removal is generically a game of strategic substitutes, which generates a free-rider problem \citep{klima2018space}. In these cases, higher mean eigenvector centrality suggests that optimal interventions may need to target either fewer objects or the same number of objects with less intensity---good outcomes for a budget-constrained planner. But when the local environment is more crowded, maneuvers may be strategic complements: if I maneuver to avoid a debris fragment, I may (perhaps unintentionally) force you to maneuver either due to the fragment's continued trajectory (which was blocked in simulations by my presence) or due to my own altered trajectory (which may now intersect yours).\footnote{I ignore the case of adversarial maneuvers, in which I maneuver with the explicit intent to force you off your path. Such actions are necessarily strategic complements.}  In these cases, higher mean eigenvector centrality suggests that optimal interventions may need to target more objects or the same number of objects with more intensity---bad outcomes for a budget-constrained planner. \\

\begin{table}[htbp]
\label{tab:centrality-models-results-eigen}
\caption{Centrality models: Eigenvector centrality as outcome}
\centering
\begin{adjustbox}{max width=1.15\textwidth}
\begin{tabular}[t]{lccccc}
\toprule
  & (1) & (2) & (3) & (4) & (5)\\
\midrule
$M_{it}$ & \num{0.095}*** & \num{0.076}*** & \num{0.063}*** & \num{0.018}*** & \\
 & (\num{0.005}) & (\num{0.005}) & (\num{0.004}) & (\num{0.004}) & \\
$CA^{C1408}_{it}$ &  & \num{0.002}*** & \num{0.002}*** & \num{0.002}*** & \num{0.002}***\\
 &  & (\num{0.000}) & (\num{0.000}) & (\num{0.000}) & \vphantom{3} (\num{0.000})\\
$CA^{Other}_{it}$ &  & \num{0.002}*** & \num{0.002}*** & \num{0.002}*** & \num{0.002}***\\
 &  & (\num{0.000}) & (\num{0.000}) & (\num{0.000}) & \vphantom{2} (\num{0.000})\\
$CA^{C1408}_{it-1}$ &  & \num{0.000} & \num{0.000} & \num{0.000}** & \num{0.000}**\\
 &  & (\num{0.000}) & (\num{0.000}) & (\num{0.000}) & \vphantom{1} (\num{0.000})\\
$CA^{Other}_{it-1}$ &  & \num{0.001}*** & \num{0.001}*** & \num{0.001}*** & \num{0.001}***\\
 &  & (\num{0.000}) & (\num{0.000}) & (\num{0.000}) & (\num{0.000})\\
$\log(E[Pc_{it}])$ &  &  & \num{0.045}*** & \num{0.004}* & \num{0.004}*\\
 &  &  & (\num{0.003}) & (\num{0.002}) & (\num{0.002})\\
$M_{it} \times$ C1408--1 &  &  &  &  & \num{0.068}***\\
 &  &  &  &  & (\num{0.014})\\
$M_{it} \times$ C1408--2 &  &  &  &  & \num{0.018}***\\
 &  &  &  &  & (\num{0.005})\\
$M_{it} \times$ C1408--3 &  &  &  &  & \num{0.016}***\\
 &  &  &  &  & (\num{0.003})\\
$M_{it} \times$ C1408--4 &  &  &  &  & \num{0.015}\\
 &  &  &  &  & (\num{0.054})\\
\midrule
Num.Obs. & \num{2729030} & \num{2513281} & \num{2513281} & \num{2513281} & \num{2513281}\\
FE: pre/post ASAT & X & X & X & X & X\\
FE: Satellite ID &  &  &  & X & X\\
FE: C1408 neighborhood & X & X & X & X & \\
\bottomrule
\multicolumn{6}{l}{\rule{0pt}{1em}* p $<$ 0.05, ** p $<$ 0.01, *** p $<$ 0.001.}\\
\multicolumn{6}{l}{\rule{0pt}{1em}Standard errors in parentheses. All standard errors are Driscoll-Kraay with $L=4$.}\\
\end{tabular}
\end{adjustbox}
\end{table}

Given results in table \ref{tab:crowdout-models-results}, it seems like on average the situation in orbit was more like a case of strategic substitutes than one of strategic complements, at least during the sample period studied here. While that does not rule out the possibility that there were short periods in specific orbital regions where strategic complementarity was more pronounced, for this to be true globally over the sample seems inconsistent with C1408 CAs crowding out Other CAs rather than crowding them in. Further research on this topic, for example utilizing structural network formation models, may shed light on conditions under which either case may be true.



\section{Conclusion}
\label{sec:conclusion}

In this paper I examine the effects of partially-automated traffic systems in LEO, focusing on the effects of close encounters between objects on orbital configurations in the aftermath of the C1408 ASAT test. Results indicate that disruptive uncontrollable objects, such as C1408 fragments, can displace close encounters between other objects. This effect is more pronounced for objects nearer, in terms of network topology, to the disruption source. Automated maneuvers can decrease collision cascade risks while amplifying potential strategic interactions between payloads. \\

I find that automated maneuvers marginally yet significantly reduce betweenness centrality, enhancing system resilience to collision cascades. Simultaneously, they marginally but significantly increase eigenvector centrality. The nature of strategic interactions---substitutes or complements---determines what this means for optimal interventions by budget-constrained planners. Since C1408 close encounters seem to displace other close encounters, maneuvers are likely strategic substitudes, implying that increased eigenvector centrality reflects reduced costs for further (optimal) interventions. \\

This paper underscores the significance of understanding network dynamics and strategic interactions for space traffic management. The findings inform the design of future space traffic systems and the impact of disruptions in other high-speed partially-automated transportation systems on traffic patterns and low-probability, high-impact failures. Further research, including optimal intervention strategies under varying network structures and scenarios, may offer valuable insights into effective space traffic management and resilience against disruptive events. As traffic systems become increasingly automated and object interactions more complex, comprehending the risks and benefits of different traffic management strategies is crucial.



\newpage
{
	\setlength{\bibsep}{3pt}
	\setstretch{1}
	\bibliography{database}
	\bibliographystyle{jpe}
}



\end{document}